\documentclass[aip,amsmath,amssymb,reprint,twocolumn]{revtex4-1}
\pdfoutput=1

\usepackage[T1]{fontenc}
\usepackage{rotating}
\usepackage{listings}
\usepackage{xcolor}
\usepackage{lipsum}
\usepackage{wrapfig}
\definecolor{codegreen}{rgb}{0,0.6,0}
\definecolor{codegray}{rgb}{0.5,0.5,0.5}
\definecolor{codepurple}{rgb}{0.58,0,0.82}
\definecolor{backcolour}{rgb}{0.95,0.95,0.92}
\lstdefinestyle{mystyle}{
    backgroundcolor=\color{backcolour},   
    commentstyle=\color{codegreen},
    keywordstyle=\color{magenta},
    numberstyle=\tiny\color{codegray},
    stringstyle=\color{codepurple},
    basicstyle=\ttfamily\small,
    breakatwhitespace=false,         
    breaklines=true,                 
    captionpos=b,                    
    keepspaces=true,                 
    numbersep=5pt,                  
    showspaces=false,                
    showstringspaces=false,
    showtabs=false,                  
    tabsize=2
}
 
\usepackage{times,amsmath}
\usepackage{color}
\usepackage{ulem}
\usepackage{graphicx}
\usepackage{bm}
\usepackage{hyperref}
\usepackage{multirow}
\usepackage{booktabs}
\hypersetup{colorlinks=true, citecolor=blue, filecolor=blue, linkcolor=blue, urlcolor=blue}
\usepackage[ruled,vlined,linesnumbered]{algorithm2e}
\DontPrintSemicolon
\SetAlgoLined
\SetInd{0.6em}{1.2em}
\SetKwInput{KwRequire}{Require}
\SetKwInput{KwEnsure}{Ensure}
\SetKwFor{For}{for}{do}{end for}
\SetKw{Return}{return}
\usepackage{tikz}
\usetikzlibrary{shapes.geometric, arrows.meta, positioning, fit, backgrounds, calc, shadows}
\usepackage{microtype}
\begin{document}

\title{Crystal Generation using the Fully Differentiable Pipeline and Latent Space Optimization}

\author{Osman Goni Ridwan}
\email{oridwan@charlotte.edu}
\affiliation{Department of Mechanical Engineering and Engineering Science, University of North Carolina at Charlotte, Charlotte, NC 28223, USA}

\author{Gilles Frapper}
\affiliation{Applied Quantum Chemistry Group, Poitiers University-CNRS, Poitiers 86073, France}

\author{Hongfei Xue}
\affiliation{Department of Computer Science, University of North Carolina at Charlotte, Charlotte, NC 28223, USA}

\author{Qiang Zhu}
\email{qzhu8@charlotte.edu}
\affiliation{Department of Mechanical Engineering and Engineering Science, University of North Carolina at Charlotte, Charlotte, NC 28223, USA}
\affiliation{North Carolina Battery Complexity, Autonomous Vehicle and Electrification (BATT CAVE) Research Center, Charlotte, NC 28223, USA}


\date{\today}

\begin{abstract}
We present a materials generation framework that couples a symmetry-conditioned variational autoencoder (CVAE) with a differentiable SO(3) power spectrum objective to steer candidates toward a specified local environment under the crystallographic constraints. In particular, we implement a fully differentiable pipeline to enable batch-wise optimization on both direct and latent crystallographic representations. Using the GPU acceleration, this implementation achieves about fivefold speed compared to our previous CPU workflow, while yielding comparable outcomes. In addition, we introduce the optimization strategy that alternatively performs optimization on the direct and latent crystal representations. This dual-level relaxation approach can effectively escape local minima defined by different objective gradients, thus increasing the success rate of generating complex structures satisfying the target local environments. This framework can be extended to systems consisting of multi-components and multi-environments, providing a scalable route to generate material structures with the target local environment.
\end{abstract}

\maketitle
\makeatletter
\setcounter{topnumber}{2}
\setcounter{bottomnumber}{2}
\setcounter{totalnumber}{4}
\renewcommand{\topfraction}{0.9}
\renewcommand{\bottomfraction}{0.8}
\renewcommand{\textfraction}{0.07}
\renewcommand{\floatpagefraction}{0.85}

\section{Introduction}
To enable rapid materials discovery prior to synthesis, it is pivotal to have a reliable and efficient crystal structure prediction (CSP) method. In the past, global optimization strategies have achieved major successes \cite{Oganov-NRM-2019}. However, they are mostly limited to small unit cells. For large unit cells and high-throughput exploration of a large compositional space, the computational cost becomes prohibitive.  Since 2018, machine-learning (ML) and artificial intelligence (AI) approaches have been increasingly applied to the CSP field. These models offer a fundamentally new approach to exploring structure space by learning data-driven representations, allowing for the efficient generation of low-energy crystal structures at a much faster rate than traditional global optimization methods. Despite the ongoing criticisms \cite{cheetham2024artificial}, the AI-based CSP approaches have been widely used to accelerate the discovery of new materials such as magnets, ferroelectrics, and thermoelectric materials \cite{merchant2023scaling, zeni2025generative}.

Recently, we introduced a symmetry-informed approach called Local Environment Geometry-Oriented Crystal Generator (\texttt{LEGO-xtal}) \cite{ridwan2025ai} to rapidly generate low-energy crystals with a target motif, using sp$^2$ carbon allotropes as the benchmark example. Building on this foundation, we report the further implementation of a GPU-enabled differentiable framework to explore alternative optimization strategies at both direct and latent structural representation space. In the practical application, we demonstrate that the dual optimization strategy can notably improve the success rate of generating low-energy new sp$^2$ carbon allotropes. This implementation is expected to pave the way for accelerated discovery of novel materials with targeted local coordination environments.

\section{Related Work}

\subsection{Global optimization based CSP methods}
Most CSP algorithms cast structure discovery as a global optimization problem on a rugged potential-energy surface. Representative methods include evolutionary algorithms~\cite{Lyakhov-CPC-2013,Lonie-CPC-2011,GASP-Python}, particle-swarm and swarm-intelligence searches~\cite{wang2012calypso}, random structure searching~\cite{Pickard-JPCM-2011}, basin-hopping~\cite{banerjee2021crystal}, and related metadynamics and genetic search variants~\cite{Zhu-CE-2012, QZhu-PRB-2015}. While these strategies have enabled many high-impact discoveries~\cite{Oganov-NRM-2019}, their practical throughput is commonly dominated by repeated local relaxations and extensive samplings. This becomes increasingly challenging as the number of degrees of freedom grows (e.g. unit cells with tens to hundreds of atoms, multiple Wyckoff sites, and broad symmetry/composition search spaces), motivating complementary data-driven generation and accelerated screening workflows.

\subsection{AI generative models for crystals}
Deep generative modeling has become popular for crystal generation, spanning Variational Auto Encoder (VAE)-based latent-variable models \cite{kingma2013auto,CDVAE}, Generative Adversarial Network (GAN)-based formulations \cite{kim2020generative}, diffusion models \cite{DiffCSP}, and autoregressive generators \cite{antunes2024crystal}. These methods offer different trade-offs between sampling cost, controllability, and output fidelity, and they form the basis for modern ML-driven CSP pipelines.
More recently, incorporating crystallographic priors is widely recognized as essential for validity and control, including space-group--conditioned diffusion \cite{diffcsp++}, Wyckoff-based generators \cite{zhu2024wycryst}, and symmetry-informed transformer architectures \cite{cao2024space}.

\subsection{Machine learning for crystal screening and relaxation}
After the generation of candidates, high-throughput workflows typically rely on fast screening models and machine-learned interatomic potentials (MLIPs) to reduce the number of structures that require expensive first-principles evaluation. For reference, the final validation of stability is commonly carried out using electronic-structure codes such as \texttt{VASP} \cite{Vasp-PRB-1996} and \texttt{Quantum Espresso} \cite{QE}. In addition, MLIPs based on message passing and E(3)-equivariant architectures provide accelerated geometry relaxation and energy ranking across large candidate pools, improving practical throughput for screening and database construction~\cite{Batatia2022mace, yang2024mattersim, wood2025family}. Other than energy-based optimization, we recently proposed the concept of ``pre-relaxation" for a rapid pre-process of massive AI-generated structures before the use of MLIPs or DFT energy-based optimizations \cite{ridwan2025ai}. 

\section{Method}

\subsection{Dataset Preparation and Augmentation}
Following our previous work on \texttt{LEGO-xtal} \cite{ridwan2025ai}, we focus on developing a framework to harness sp$^2$ carbon allotropes. To begin, we construct the training set from the SACADA database~\cite{hoffmann2016homo}, which contains 154 experimentally known or hypothetical sp$^2$-bonded carbon allotropes. 
Here we adopt a compact tabular representation that exploits space-group symmetry to reduce dimensionality while remaining fully reconstructible. Each structure is encoded as a fixed-length vector comprising: (i) the space-group number (\texttt{spg}); (ii) the six lattice parameters $(a,b,c,\alpha,\beta,\gamma)$; and (iii) Wyckoff-site positions (WPs), each specified by a Wyckoff position index and the fractional coordinates of the corresponding representative atom in the asymmetric unit.
By default, crystal structures are presented in the highest possible symmetry group. To increase diversity while preserving crystallographic validity, we applied subgroup augmentation~\cite{zhu2022symmetry}: for each parent structure, we enumerate possible subgroup combinations and the symmetry operations of the Wyckoff sites. This procedure yields 63 \,115 structures spanning a broad range of symmetry combinations. 

\subsection{Conditional VAE}
In the \texttt{LEGO-xtal} approach \cite{ridwan2025ai}, we originally employed the standard VAE model to handle both continuous variables (lattice parameters and fractional coordinates) and discrete variables (space group and Wyckoff indices). In this work, we instead adopt a two-stage approach: in the first stage, we have trained the VAE or GAN~\cite{goodfellow2014generative} model on the discrete part of the dataset (space group and Wyckoff indices) to generate valid symmetry combinations. The target of the first stage model is to learn the distribution of valid (spg, wps) combinations from the training set and generate new combinations that will be used as conditions for the second stage model. In the second stage, we train a Conditional Variational Autoencoder (CVAE)~\cite{harvey2021conditional} architecture, in which the generative process is explicitly conditioned on discrete crystallographic symmetry information. This design enables targeted exploration and optimization of the latent space while keeping the discrete symmetry fixed, allowing the continuous degrees of freedom (cell parameters and Wyckoff fractional coordinates) to adapt toward a desired local environment.

The continuous crystallographic features $X$ are transformed using a Gaussian Mixture Model (GMM)~\cite{xuan2001algorithms} cluster-based normalization scheme. Each scalar variable is encoded as a $K$-component categorical assignment together with a standardized continuous feature, yielding a transformed representation $X^\prime$. 
This GMM-based encoding helps to handle the strong multi-modality in crystallographic parameters and makes the representation smoother.
The discrete symmetry information, consisting of the space-group index and Wyckoff site labels, is converted into a one-hot condition vector $C$. Here, $X$ denotes ground-truth input features, while $\tilde{X}$ denotes the corresponding features generated by the model.

The encoder network takes $X^\prime$ as input and outputs the parameters $(\mu, \log\sigma^2)$ of a Gaussian latent distribution. Latent vector $Z$ is sampled using the reparameterization trick~\cite{kingma2013auto}. 
In parallel, the condition vector $C$ is processed by a multilayer perceptron to produce a 128-dimensional condition embedding $e_c$. The sampled latent variable $Z$ is concatenated with $e_c$ and passed to the decoder, which reconstructs the GMM-transformed features $\tilde{X'}$.
The training objective consists of a combination of reconstruction and regularization terms. Cross-entropy loss ($L_{\mathrm{CL}}$) is applied to the discrete GMM component assignments, while the continuous features are optimized using negative log-likelihood ($L_{\mathrm{NL}}$). These reconstruction losses are combined with a Kullback--Leibler divergence regularization term ($L_{\mathrm{KL}}$) to enforce a smooth latent prior and promote generalizable latent representations.

During inference, a random noise $z \sim \mathcal{N}(0, I)$ is sampled from the standard Gaussian prior, and the condition embedding $e_c$ is computed for a specified symmetry configuration $C$. The decoder generates the GMM-transformed features $\tilde{X'}$, which are mapped back to physical crystallographic parameters $\tilde{X}$ via the inverse GMM transform, yielding structures that satisfy the imposed space-group and Wyckoff position constraints.

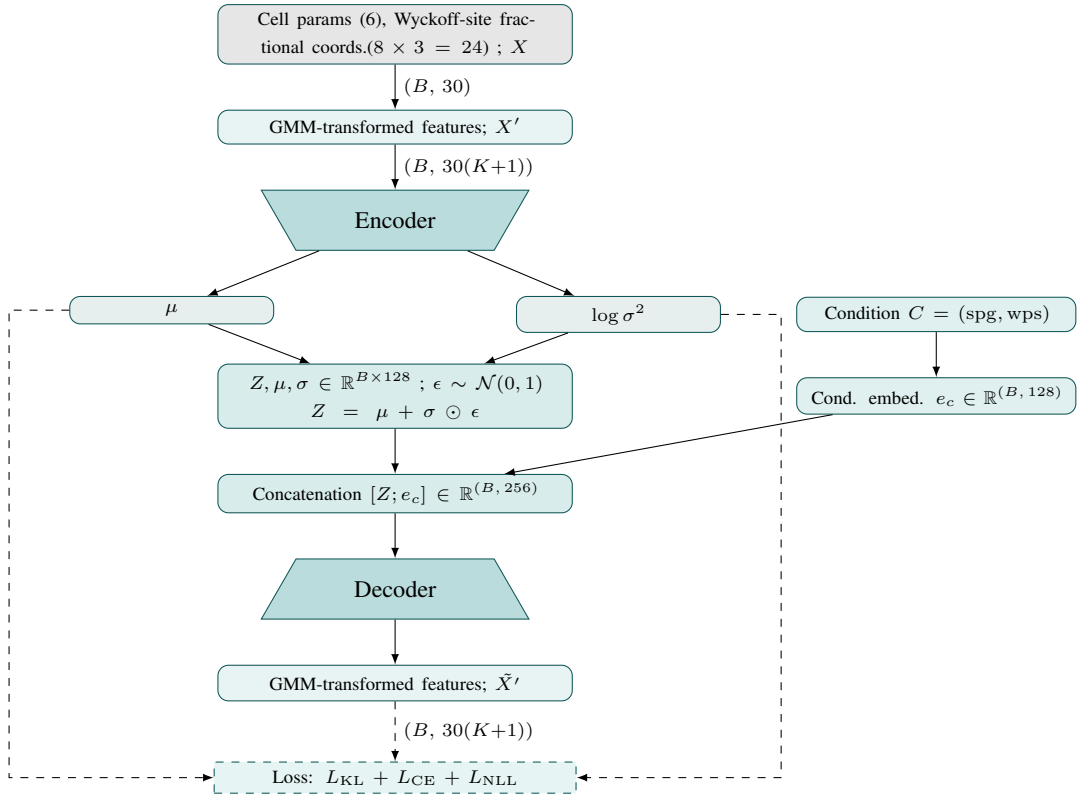
\begin{figure*}[t]
\centering
\begin{tikzpicture}[>=latex, node distance=6mm, every node/.style={font=\small}]
  \definecolor{PaperTeal}{HTML}{66B2B2}      
  \definecolor{PaperTealDark}{HTML}{105656}  
  \definecolor{PaperTealPale}{HTML}{E6F4F4}  
  \definecolor{PaperGray}{HTML}{E6E6E6}

  \tikzset{
    input/.style={draw=PaperTealDark, rounded corners, fill=PaperGray, align=center, text width=4.5cm},
    process/.style={draw=PaperTealDark, rounded corners, fill=PaperTeal!15, align=center, text width=4.5cm},
    encblock/.style={draw=PaperTealDark, trapezium, trapezium left angle=120, trapezium right angle=120,
      fill=PaperTeal!45, align=center, minimum height=8mm, minimum width=12mm, text width=12mm},
    decblock/.style={draw=PaperTealDark, trapezium, trapezium left angle=120, trapezium right angle=120, shape border rotate=180,
      fill=PaperTeal!45, align=center, minimum height=8mm, minimum width=12mm, text width=12mm},
    latent/.style={draw=PaperTealDark, rounded corners, fill=PaperTealDark!10, align=center, text width=2.5cm},
    cond/.style={draw=PaperTealDark, rounded corners, fill=PaperTeal!20, align=center, text width=3.5cm},
    merge/.style={draw=PaperTealDark, rounded corners, fill=PaperTeal!25, align=center, text width=4.5cm},
    lossbox/.style={draw=PaperTealDark, dashed, fill=PaperTealPale, align=center, text width=4.6cm}
  }

  \node[input] (x) {\scriptsize Cell params (6), Wyckoff-site fractional coords.($8 \times 3 = 24$) ; $X$};

  \node[process, below=of x] (gmm)
    {\scriptsize GMM-transformed features; $X'$};
  \node[encblock, below=of gmm] (enc)
    {Encoder};
  
  \node[latent, below left=6mm and 12mm of enc] (mu) {\scriptsize $\mu$};
  \node[latent, below right=6mm and 12mm of enc] (logvar) {\scriptsize $\log\sigma^2$};
  \node[merge, below=15mm of enc] (z)
    {\scriptsize $Z,\mu,\sigma\in\mathbb{R}^{B\times 128}$ ; $\epsilon\sim\mathcal{N}(0,1)$\\$Z = \mu + \sigma\odot\epsilon$};
  
  \node[cond, right=10mm of logvar] (cond)
    {\scriptsize Condition $C=(\mathrm{spg},\mathrm{wps})$};
  \node[cond, below=of cond] (cemb)
    {\scriptsize Cond. embed. $e_c\in\mathbb{R}^{(B,\,128)}$};
  
  \node[merge, below=of z] (concat)
    {\scriptsize Concatenation $[Z;e_c]\in\mathbb{R}^{(B,\,256)}$};
  
  \node[decblock, below=of concat] (dec)
    {Decoder};
  \node[process, below=of dec] (out)    
    {\scriptsize GMM-transformed features; $\Tilde{X'}$};
  
  \node[lossbox, below=8mm of out] (loss)
    {\scriptsize Loss: $L_{\mathrm{KL}} + L_{\mathrm{CE}} + L_{\mathrm{NLL}}$};

  \draw[->] (x) -- node[right, font=\scriptsize] {$(B,\,30)$} (gmm);
  \draw[->] (gmm) -- node[right, font=\scriptsize] {$(B,\,30(K{+}1))$} (enc);
  \draw[->] (enc) -- (mu);
  \draw[->] (enc) -- (logvar);
  \draw[->] (mu) -- (z);
  \draw[->] (logvar) -- (z);
  \draw[->] (z) -- (concat);
  
  \draw[->] (cond) -- (cemb);
  \draw[->] (cemb) -- (concat);
  
  \draw[->] (concat) -- (dec);
  \draw[->] (dec) -- (out);
  
  \draw[->, dashed] (out) -- node[right, font=\scriptsize] {$(B,\,30(K{+}1))$} (loss);     
  \draw[->, dashed] (mu.west) -| ++(-8mm,0) |- (loss.west);
  \draw[->, dashed] (logvar.east) -| ++(8mm,0) |- (loss.east);

\end{tikzpicture}
\caption{Architecture of the Conditional VAE with DiffGMM data transformation. The encoder processes GMM-transformed continuous features to produce latent variables $\mu$ and $\log\sigma^2$, while discrete conditions (space group and Wyckoff positions) are embedded separately and concatenated with the sampled latent $z$ before decoding.}
\label{fig:cvae_architecture}
\end{figure*}

\subsection{SO(3) Descriptor and Environmental Loss}
To guide crystal structure optimization toward desired bonding motifs, we adopt the SO(3) power spectrum descriptor \cite{Bartok-PRB-2013} to encode both radial and angular information of an atom's local neighborhood in a rotation-invariant form. 
In practice, the descriptor ($P$) is computed by expanding a Gaussian-smeared neighbor density around each atom in a combined radial and spherical harmonic basis, followed by forming rotation-invariant power spectrum components. To describe the local sp$^2$ bonding motif as found in the graphite structure with coordination number $\mathrm{CN} = 3$, we compute $P$ with several hyper-parameters to truncate the spread of basis, including (1) $n_{\mathrm{max}} = 2$ controlling the radial resolution, (2) $\ell_{\mathrm{max}} = 4$ for the maximum angular momentum, (3) $r_{\mathrm{cut}} = 2.0\,\text{\AA}$ for the local neighborhood radius considered around each atom, (4) $\alpha = 1.5\,\text{\AA}$ for the Gaussian-smeared width. Further mathematical details are provided in the online code and previous literature \cite{yanxon2020pyxtalff}.

Each generated crystal is evaluated by computing $P$ for all Wyckoff sites and comparing them to the reference descriptor extracted from graphite. The per-structure loss is defined as the mean-squared deviation between the computed and reference descriptors,

\begin{equation}
\label{eq:desc_loss}
\ell_i = \frac{1}{W_i L} \sum_{j=1}^{W_i} \sum_{k=1}^{L} (P_{ijk} - P_{\mathrm{ref},k})^2,
\end{equation}

where $i$ is the index for structure, $j$ is the index for Wyckoff site, and  $k$ indexes the SO(3) descriptor components. Correspondingly, $W_i$ is the number of Wyckoff sites in structure $i$, $P_{ijk}$ denotes the descriptor component $k$ for the wyckoff site $j$ in structure $i$, and $P_{\mathrm{ref}}$ is the reference descriptor in the vector format.

\subsection{Descriptor based Optimization}
\begin{figure*}[t]
  \centering
  \includegraphics[width=\textwidth]{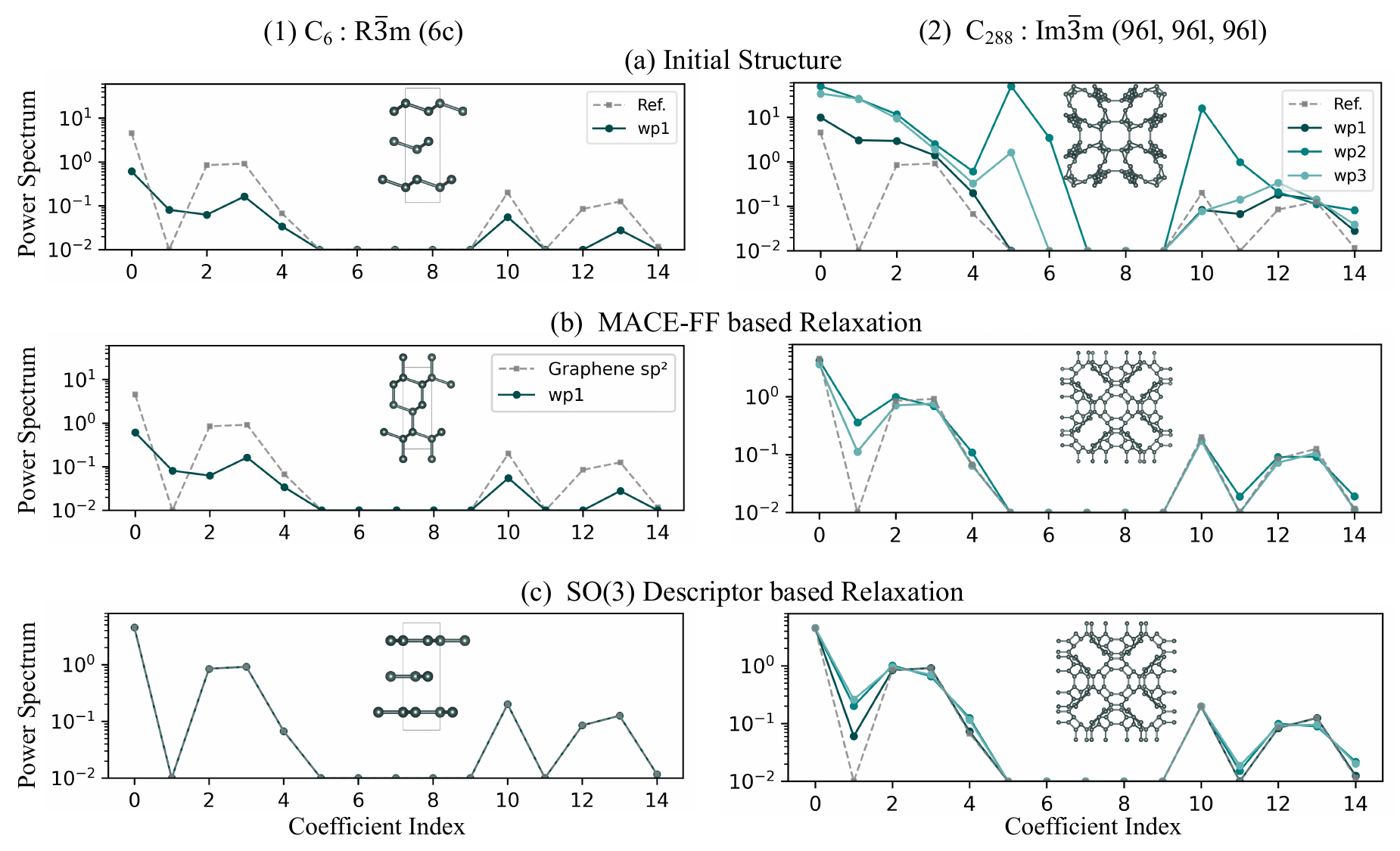}
    \vspace{-5mm}
  \caption{SO(3) power spectrum comparison for CVAE-generated carbon structures before and after refinement. Shown are the initial structures, results after energy-based relaxation using MACE-FF, and results after SO(3) descriptor-based optimization, together with the graphite sp$^2$ reference (gray dashed line).}
  \label{fig:descriptor}
\end{figure*}

In the recent work \cite{ridwan2025ai}, we have implemented an optimization routine to minimize the mean-squared deviation from the target local-environment fingerprint defined in Eq.~\ref{eq:desc_loss}, by taking the advantage of \textit{scipy.minimize} library. In Fig.~\ref{fig:descriptor}, we have shown two examples: (1) a small C$_6$ structure (space group 166 with one $6c$ Wyckoff site) and (2) a large C$_{288}$ structure (space group 229 with three $96l$ sites) to illustrate a common failure mode of energy-only relaxation and the role of descriptor guidance.
In both cases, the initial generated structures exhibit substantial deviations from the graphite reference power spectrum (gray dashed lines), indicating that their local atomic environments differ significantly from the target motif.

We compare the descriptor based minimization approach with the conventional energy-based relaxation using the \textit{MACE-mp-0} interatomic potential ~\cite{batatia2025foundation}. For the C$_6$ (Fig~\ref{fig:descriptor}-(1)) structure, the MACE energy-based relaxation converges to a lower-energy configuration with the sp$^3$ bonding in a tetrahedral coordination. In contrast, descriptor-based optimization yields a final structure with the desired sp$^2$ bonding. This highlights that energy minimization alone may not guarantee convergence toward a desired bonding motif when multiple local energy minima are present.
For the larger C$_{288}$ (Fig~\ref{fig:descriptor}-(2)) structure, MACE-based relaxation lowers the energy to $-8.518$~eV/atom but retains noticeable discrepancies as compared to the reference environment. Descriptor-based optimization yields nearly sp$^2$ bonding arrangement across all Wyckoff sites, while maintaining a comparable final energy of $-8.480$~eV/atom. Importantly, the descriptor-based optimization reaches this state significantly faster ($27.8$~s vs. $6.2$~min on the same CPU hardware). If we further perform MACE relaxation on this relaxed structure, it would become essentially the identical structure after only a few iterations. 
Therefore, the descriptor-based optimization can serve as a fast pre-relaxation tool for two purposes: (1) to steer generative model outputs toward a specified local bonding motif, and (2) to reduce the computational cost for structural relaxation.

In our previous \texttt{LEGO-xtal} framework \cite{ridwan2025ai}, we sequentially transformed each structure to Cartesian coordinates, computed the neighbor lists, radial density functions with spherical Bessel and harmonic projections, and calculated the loss function on the CPU device. 
In this approach, the optimization relied on either derivative-free updates (e.g.\ Nelder--Mead~\cite{nelder1965simplex}) or gradient-based routines that required finite-difference gradients in our setting (e.g.\ L-BFGS-B~\cite{zhu1997algorithm}). While this strategy is sufficient for handling datasets smaller than $500$\,K with small unit cells, it becomes a computational bottleneck when processing large candidate pools from generative models. In addition, the success rate of converging to target local environments remains limited, particularly for complex structures with multiple Wyckoff sites where the descriptor loss landscape exhibits multiple local minima. Hence, we aim to overcome these limitations by leveraging GPU acceleration and automatic differentiation to enable batch-wise optimization, combined with a dual-level refinement strategy that exploits both the representation space and the learned CVAE latent space.

\subsection{GPU-accelerated Optimization on the Representation-Space}

To start, we seek to speed up the optimization by leveraging fast automatic differentiation on the GPU platform~\cite{paszke2017automatic}.
The key challenge is enabling automatic differentiation to compute gradients through the entire SO(3) descriptor calculation for thousands of structures in parallel.
 The workflow is summarized in Algorithm \ref{alg:batch_relax}. First, from the tabular representation, we identify the minimal set of free variables (reduced parameters) for each structure based on its space group symmetry constraints. We refer to this symmetry-reduced parameterization as the representation space. For instance, cubic systems require only one lattice parameter instead of six, and special Wyckoff positions fix certain fractional coordinates. We encode these free variables as a learnable tensor $\mathbf{R} \in \mathbb{R}^{B\times N}$ with padding and masking for unused entries (where $B$ is the batch size and $N$ is the length of continuous variables), and the descriptor loss is evaluated for all structures in parallel. Automatic differentiation then provides gradients of the descriptor loss with respect to $\mathbf{R}$, enabling efficient gradient-based updates at scale. We sort structures by space group before batching. Structures sharing the same space group have similar Wyckoff-site options and free-variable layouts, which reduces padding in the fixed-length representation and improves GPU memory efficiency.
This strategy enables processing of large batches (e.g., $1000$ structures per batch) on modern GPUs, delivering substantial wall-clock speedup compared to sequential CPU processing.

\begin{algorithm}[htbp]
\caption{Batch-wise optimization on the continuous crystal representation $\mathbf{R}$.}
\label{alg:batch_relax}
\KwRequire{batch rows $\tilde{X}$, WP constructor, calculator $f$, reference descriptor $p_{\rm ref}$, steps $T$}
\KwEnsure{optimized $\mathbf{R}$ and final losses $\{\ell_i\}$}
\BlankLine

Encode batch: $(\mathrm{spg},\ \mathrm{wps},\ \mathbf{R}) \gets \tilde{X}$\;
Initialize $\mathbf{R}$ as learnable and set up a gradient-based optimizer\;
Initialize per-sample state (best loss, plateau counter, and a scaling factor)\;
\BlankLine

\For{$t\gets 1$ \KwTo $T$}{
  Reconstruct batch geometry from $(\mathrm{spg},\ \mathrm{wps},\ \mathbf{R})$ using WP constructor\;
  Compute SO(3) descriptors for all structures in the batch: $P \gets f(\cdot)$\;
  Compute per-sample losses $\{\ell_i\}$ using Eq.~\ref{eq:desc_loss} (masking padded/unused Wyckoff entries)\;
  Backpropagate $\ell = \sum_i \ell_i$ to obtain gradients with respect to $\mathbf{R}$\;
  Apply per-sample gradient conditioning (scaling on plateau and gradient clipping)\;
  Optimizer update on $\mathbf{R}$ and clamp normalized variables to $[0,1]$\;
}
\Return optimized $\mathbf{R}$ and final losses $\{\ell_i\}$\;
\end{algorithm}

In the code implementation, we precomputed index mapping tensors for each Wyckoff sites to enable vectorized structure reconstruction on GPU. For each batch, we store generator coordinates for the occupied Wyckoff sites along with mapping tensors that (i) associate each free variable in $\mathbf{R}$ with the corresponding coordinates and (ii) enumerate the symmetry operations needed to expand each generator into the full atomic basis. Using $(\mathrm{spg}, \mathrm{wps}, \mathbf{R})$ and these mapping tensors, we reconstruct the batched structures and compute the forward loss function as a single batched tensor program on GPU. Automatic differentiation then provides gradients with respect to $\mathbf{R}$ in the backward mode, enabling efficient gradient-based optimization for the whole batch structures.

In practice, structures that are already close to the target (small loss) in a batch may stop improving because their gradients are dominated by a few hard samples with large loss. To mitigate this issue, we employ the AdamW optimizer~\cite{loshchilov2018decoupled} with per-structure adaptive conditioning: gradients are clipped independently for each sample, and the effective learning rate is adaptively reduced when a sample's loss plateaus. This approach enables independent optimization of heterogeneous structures within a batch and prevents difficult cases from dominating the collective update, and improves convergence across all samples.

\subsection{GPU-accelerated Optimization on the Latent Space and Dual-Level Refinement}
After the use of Algorithm ~\ref{alg:batch_relax}, a significant fraction of samples may still fail to reach the desired local environments. In our previous framework, these structures were simply discarded. From an optimization perspective, many structures can get trapped in local minima defined by the descriptor loss landscape, making it challenging to achieve the target motif through local perturbations of the free variables $\mathbf{R}$. During CVAE training, the model not only generates final structures but also learns a latent space $Z$ that captures correlations among crystallographic parameters under fixed symmetry conditions. Compared to the direct representation space, this latent space may offer a smoother optimization landscape that facilitates escaping local minima.
In this latent space, we can explore coordinated changes across multiple degrees of freedom simultaneously, potentially escaping local minima that are difficult to overcome through direct perturbations in the representation space.

Hence, we propose another optimization operated on the CVAE latent space $Z$ while keeping the discrete symmetry condition fixed (see Algorithm \ref{alg:latent_opt}). Instead of directly perturbing the free variables $\mathbf{R}$ (lattice degrees of freedom and Wyckoff free coordinates), we modify the $Z$ variables from the decoder $D_\theta(Z, C)$ to generates new continuous crystal parameters under a fixed $(\mathrm{spg}, \mathrm{wps})$ conditions. Because the decoder is trained to model correlations among these parameters, small changes in $Z$ can induce collective adjustments across lattice and Wyckoff-site coordinates, rather than changing each variable independently. This coupling provides an effective mechanism to move candidates between distinct structural modes that may be difficult to reach using local steps in the $\mathbf{R}$ space. 

%
\begin{algorithm}[t]
\caption{Batch-wise optimization on the latent space ($Z$).}
\label{alg:latent_opt}
\KwRequire{batch conditions $\{C\}$, CVAE decoder, WP constructor, calculator $f$, reference descriptor $p_{\rm ref}$, steps $T$}
\KwEnsure{optimized latent variables $\{Z\}$ and decoded $\tilde{X}$}

\BlankLine
Initialize latent variables $\{Z\}\in\mathbb{R}^{B\times 128}$

\For{$t\gets 1$ \KwTo $T$}{
  Decode $(Z,C)$ to $\tilde{X}'$; inverse GMM to $\tilde{X}$; reconstruct $(spg,wps,\mathbf{R}))$;
  Reconstruct batched geometries via WP constructor\;
  Compute descriptors for the batch: $P \gets f(\cdot)$\;
  Compute per-sample losses $\{\ell_i\}$ using Eq.~\ref{eq:desc_loss}\;
  Backpropagate the summed loss to obtain gradients w.r.t. $\{Z\}$\;
  Apply the same per-sample conditioning as Alg.~\ref{alg:batch_relax} and update $\{Z\}$\;
}
\Return optimized $\{Z\}$ and decoded $\tilde{X}$;
\end{algorithm}

\begin{figure}[t]
\centering
\begin{tikzpicture}[>=latex, node distance=5mm, every node/.style={font=\footnotesize, align=center}, scale=0.85, transform shape]
  \definecolor{PaperTeal}{HTML}{66B2B2}      
  \definecolor{PaperTealDark}{HTML}{105656}  
  \definecolor{PaperTealPale}{HTML}{E6F4F4}  
  \definecolor{PaperGray}{HTML}{E6E6E6}

  \tikzset{
    startstop/.style={draw=PaperTealDark, rounded corners, fill=PaperTeal!25, minimum width=3.5cm},
    process/.style={draw=PaperTealDark, rounded corners, fill=PaperTeal!15, minimum width=3.5cm},
    opt/.style={draw=PaperTealDark, rounded corners, fill=PaperTealDark!10, minimum width=3.5cm},
    decision/.style={draw=PaperTealDark, diamond, aspect=2, fill=PaperTealPale},
    neutral/.style={draw=PaperTealDark, rounded corners, fill=PaperGray},
    iter/.style={draw=PaperTealDark, diamond, aspect=2, fill=PaperTeal!18},
    update/.style={draw=PaperTealDark, rounded corners, fill=PaperTeal!18},
    validset/.style={draw=PaperTealDark, rounded corners, fill=PaperTeal!25}
  }
  
  \node[startstop] (start) 
    {Generate $N=10^5$ candidates};
  
  \node[process, below=of start] (sample) 
    {Sample \& Decode};
  
  \node[opt, below=of sample] (repopt1) 
    {Rep-Opt (Alg.~\ref{alg:batch_relax})};
  
  \node[decision, below=of repopt1] (validate1) 
    {Valid (CN=3)?};

  \node[neutral, left=12mm of validate1] (invalid1) 
    {Set $\mathcal{W}$ (invalid indices)};
  
  \node[iter, below=6mm of invalid1] (iter) 
    {$r < I$?};
  
  \node[opt, below=of iter, minimum width=2.5cm] (latopt) 
    {Latent-Opt (Alg.~\ref{alg:latent_opt})};
  
  \node[opt, below=of latopt, minimum width=2.5cm] (reopt) 
    {Rep-Opt (Alg.~\ref{alg:batch_relax})};
  
  \node[decision, below=of reopt] (validate2) 
    {Valid (CN = 3)?};

  \node[validset] (valid1) at ($(validate1.center |- validate2.center)$)
    {$\mathcal{D}_{\rm valid}$};
  \node[update, below=6mm of validate2] (continue) 
    {Update $\mathcal{W}$};

  \node[startstop, below=6mm of continue, minimum width=3cm] (output) 
    {Filter \& Return};

  \draw[->] (start) -- (sample);
  \draw[->] (sample) -- (repopt1);
  \draw[->] (repopt1) -- (validate1);
  \draw[->] (validate1.south) -- node[midway, right, font=\scriptsize] {Yes} (valid1.north);
  \draw[->] (validate1.west)  -- node[above, font=\scriptsize] {No}  (invalid1.east);
  
  \draw[->] (invalid1) -- (iter);
  \draw[->] (iter) -- node[right, font=\scriptsize] {Process $\mathcal{I}$} (latopt);
  \draw[->] (latopt) -- (reopt);
  \draw[->] (reopt) -- (validate2);
  
  \draw[->] (validate2.east) -- node[above, font=\scriptsize] {Yes} (valid1.west);
  \draw[->] (validate2) -- node[left, font=\scriptsize] {No} (continue);
  
  \draw[->] (continue.west) -- ++(-4cm, 0) |- (iter.west);
  
  \draw[->] (valid1.south) |- (output.east);
  
\end{tikzpicture}
\caption{Iterative dual-level refinement workflow. Structures satisfying the target $\mathrm{CN}=3$ motif are collected in $\mathcal{D}_{\rm valid}$; remaining samples ($\mathcal{W}$) undergo latent refinement followed by representation-space re-optimization for up to $I$ rounds.}
\label{fig:iterative_refinement}
\end{figure}
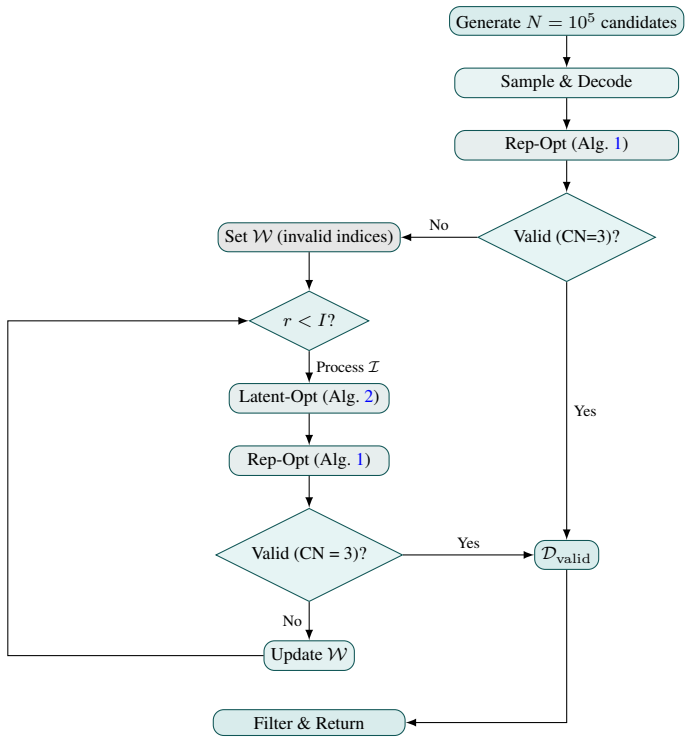

Because latent-space optimization involves decoder evaluation and backpropagation through the learned model, it is computationally more expensive than direct representation-space optimization. Consequently, we employ latent refinement as a targeted recovery step, applied only to structures that fail to reach the target local environment after optimization on the direct representation space. In practice, we implement an iterative dual-level workflow that alternates between fast representation optimization and latent-space refinement for the remaining invalid samples (see Fig.~\ref{fig:iterative_refinement}).

In Run~1, direct representation optimization is applied to all $N$ candidates, and structures satisfying the $\mathrm{CN}=3$ criterion are collected into $\mathcal{D}_{\rm valid}$. The remaining structures form the invalid set $\mathcal{I}_1$. For subsequent rounds $r=2$ or higher, we perform latent-space refinement on all samples in $\mathcal{I}_{r-1}$ while keeping their discrete symmetry conditions $C$ fixed. The refined latent variables are decoded to generate updated continuous parameters, which are then optimized via direct representation refinement. The new valid structures are added to $\mathcal{D}_{\rm valid}$, and the remaining invalid structures constitute $\mathcal{I}_r$. At each round, latent variables are reinitialized, while the symmetry condition $C$ and the set of invalid sample indices are preserved for the next iteration. In practice, we find that $2$--$3$ rounds are sufficient to recover most invalid samples, balancing computational cost and yield.

\section{Results}
We trained both a baseline VAE and a CVAE model using the same 63,115 sp$^2$ carbon data for 250 epochs with a batch size of 2048 on a single NVIDIA H100 GPU. The encoder and decoder architectures comprised $2\times1024$ hidden layers, and we used loss weights of $1{:}2{:}0.1$ for the KL, CL, and NL terms respectively, with a latent dimension of 128. After training, we generated $100$k candidates from each model for evaluation. For the CVAE, we reused the same condition data $C=(\mathrm{spg},\mathrm{wps})$ when sampling, ensuring both models were evaluated under identical symmetry constraints. On the generated dataset, we applied the descriptor-based optimization and a post-processing and screening pipeline to construct the final database. First, we retain only those structures that satisfy the target sp$^2$ coordination criterion ($\mathrm{CN} = 3$), yielding $N_{\rm valid}$ candidates. Next, we characterize each candidate using \texttt{CrystalNets.jl}~\cite{zoubritzky2022crystalnets} to determine its topological net label and structural dimensionality. We then remove duplicates using this topology--dimensionality signature, which provides an efficient first-pass uniqueness filter.
Furthermore, we relaxed the remaining candidates with the MACE potential~\cite{Batatia2022mace} and computed their energies. After relaxation, we performed a second uniqueness check using pymatgen's \texttt{StructureMatcher}~\cite{ong2013python} with the following matching tolerances: a maximum site distance of $0.3$~\AA, a maximum relative lattice parameter difference of $20\%$, and a maximum lattice angle difference of $5^\circ$. The resulting set, $N_{\rm final}$, consists of structures that are following target local motif, unique, and energetically screened. 

In the following section, we assess model performance using three metrics, including (1) $N_{\rm valid}$: the number of candidates that satisfy the $3$-coordination requirement optimization; (2) $N_{\rm unique}$: the number of unique structures after final MACE relaxation and removing duplicates; and (3)
$N_{\rm low\_E}$: the number of low energy unique structures within $0.55$~eV/atom of the reference graphite.

\subsection{The Performance of CVAE Generation} 
\begin{figure*}[t] 
  \centering 
  \begin{minipage}[t]{0.39\linewidth}\vspace{9pt} 
    \centering 
    \small
    \setlength{\tabcolsep}{10pt} 
    \renewcommand{\arraystretch}{1.5}
    \begin{tabular}{l r r} 
      \toprule 
      & Base VAE & CVAE\\  
      \midrule 
      $N_{\rm total}$      & 100000 & 100000 \\ 
      $N_{\rm valid\_env}$ & 18248  & 23372  \\ 
      $N_{\rm unique}$     & 4878   & 5695   \\
      $N_{\rm low\_E}$     & 508    & 571    \\ 
      \bottomrule 
    \end{tabular}
  \end{minipage}
  \hfill 
  \begin{minipage}[t]{0.60\linewidth}\vspace{0pt} 
    \centering 
    \includegraphics[width=\linewidth]{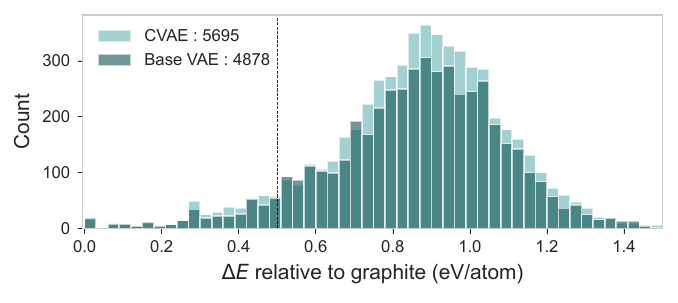} 
  \end{minipage}
  \caption{\textbf{Baseline generation comparison between a VAE and a CVAE.} Both models generate 100k candidates and are evaluated with the same post-processing pipeline. \textit{Left:} Summary counts of valid structures satisfying the target environment ($N_{\rm valid\_env}$), unique structures ($N_{\rm unique}$), and low-energy candidates ($N_{\rm low\_E}$). \textit{Right:} MACE-relaxed energy distributions (eV/atom) for the screened unique structures, shown as counts with identical binning.}
  \label{fig:cvae}
\end{figure*}

\begin{figure*}[htbp] 
  \centering 
  \begin{minipage}[t]{0.40\linewidth}\vspace{9pt} 
    \centering 
    \setlength{\tabcolsep}{0pt} 
    \renewcommand{\arraystretch}{1.2}
    \begin{tabular}{@{}p{0.34\linewidth}p{0.33\linewidth}p{0.33\linewidth}@{}}
      \toprule
      & CPU-SciPy & GPU-PyTorch \\
      \midrule
      Device & AMD 96-core & NVIDIA H100 \\
      Optimization & L-BFGS-B & AdamW \\
      Gradient & Numerical & Autograd \\
      Parallelization & Multiprocess & GPU Batch  \\
      Time (/$1$k) & $\sim$5 min & $\sim$1 min \\
      $N_{\rm valid\_env}$ & 18,248 & 12,874 \\
      $N_{\rm unique}$ & 4,878 & 4,817 \\
      $N_{\rm low\_E}$ & 508 & 453 \\
      \bottomrule
    \end{tabular}
  \end{minipage}
  \hfill 
  \begin{minipage}[t]{0.59\linewidth}\vspace{3pt}
    \centering 
    \includegraphics[width=\linewidth]{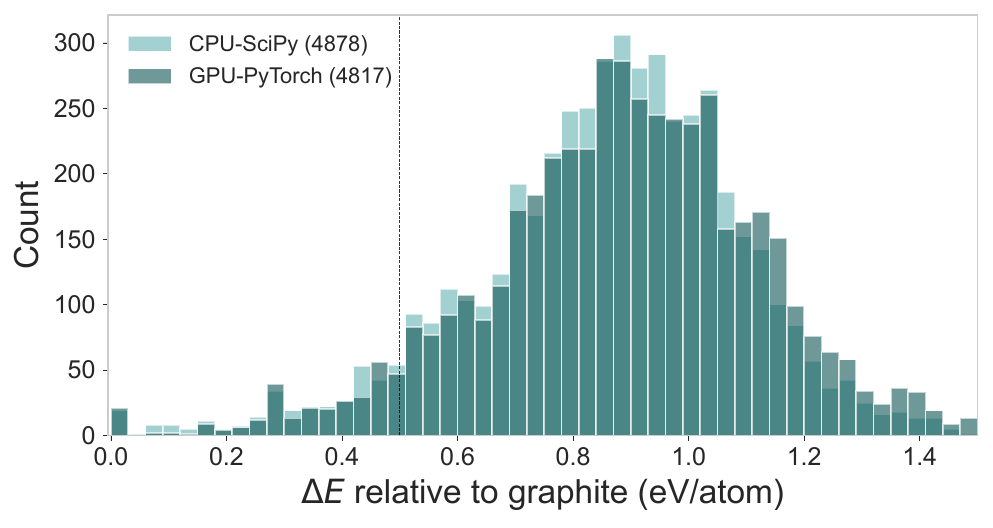} 
  \end{minipage}
  \caption{\textbf{GPU acceleration of direct representation optimization.} \textit{Left:} CPU--SciPy (prior LEGOxtal implementation) versus GPU--PyTorch (this work), including device, optimizer, batching strategy, throughput (time per 1k candidates), and post-screening counts. \textit{Right:} MACE-relaxed energy histograms (eV/atom) for the screened unique structures, shown as counts with identical binning.}
  \label{fig:torch_vs_scipy}
\end{figure*}

We compare the CVAE against the baseline VAE using the same post-processing pipeline. For a  fair comparison, we generate $100$k candidates from each model and reuse the same condition multiset $C=(\mathrm{spg},\mathrm{wps})$ when sampling the CVAE, so both models are evaluated under identical symmetry constraints.

Fig.~\ref{fig:cvae} shows that conditioning improves downstream yield. Relative to the baseline VAE, the CVAE produces more valid ($\mathrm{CN} = 3$) structures ($23{,}372$ vs $18{,}248$), retains more unique candidates after the removal of duplicates ($5{,}695$ vs $4{,}878$), and yields more low-energy structures after MACE screening ($571$ vs $508$). The MACE energy histograms for the screened unique structures overlap strongly across the full range, suggesting that the two models produce broadly similar energy profiles after screening. 

\subsection{The Performance of GPU-based Batch-wise Optimization on the Direct Representations}
We next benchmark the throughput of the proposed GPU-batched descriptor based optimization in the representation space $R$ against our previous CPU-based implementation in \texttt{LEGO-xtal}. As shown in Fig.~\ref{fig:torch_vs_scipy}, the PyTorch-GPU pipeline achieves approximately a $5\times$ reduction in wall-clock time ($\sim$5\,min to $\sim$1\,min per $1$k candidates), enabling high-throughput descriptor-guided refinement for large generative candidate pools.

Furthermore, we compare the evaluation metrics to ensure that the GPU implementation yields comparable screening outcomes. The number of unique structures is nearly identical between the two implementations ($N_{\rm unique}=4878$ for CPU--SciPy versus $4817$ for GPU--PyTorch), and the corresponding MACE energy histograms show strong overlap across the full range. There is, however, a notable difference in $N_{\rm valid}$ ($18{,}248$ for CPU--SciPy versus $12{,}874$ for GPU--PyTorch). This is likely due to the choice of optimization algorithms: our CPU runs use SciPy's L-BFGS-B with the estimated finite-difference gradients, while the GPU implementation employs batched Adam updates with automatic differentiation. Since the final sets of unique screened structures and their energy distributions are highly similar, this difference is acceptable for our purposes. Overall, the results demonstrate that our GPU implementation substantially improves throughput while maintaining a comparable post-screening energy profile.

\begin{figure*}[htbp] 
  \centering 
  \begin{minipage}[t]{0.4\linewidth}\vspace{0pt} 
    \centering 
    \setlength{\tabcolsep}{0pt} 
    \vspace{4mm}\renewcommand{\arraystretch}{1.6}
    \begin{tabular}{@{}p{0.12\linewidth}p{0.25\linewidth}p{0.20\linewidth}p{0.20\linewidth}p{0.21\linewidth}@{}}
      \toprule
      Run & N$_\text{valid\_env}$ & N$_\text{unique}$ & N$_\text{low\_E}$ & Time
      (hrs) \\
      \midrule
      1   & 15\,076 & 4\,806  & 412 & 1.66 \\
      1--2 & 29\,837 & 7\,581  & 675 & 4.36 \\
      1--3 & 40\,426 & 9\,647 & 872 & 6.66 \\
      1--4 & 48\,308 & 11\,546 & 1\,014 & 8.56 \\
      \bottomrule
    \end{tabular}
  \end{minipage}
  \hfill 
  \begin{minipage}[t]{0.57\linewidth}\vspace{3pt} 
    \centering 
    \includegraphics[width=\linewidth]{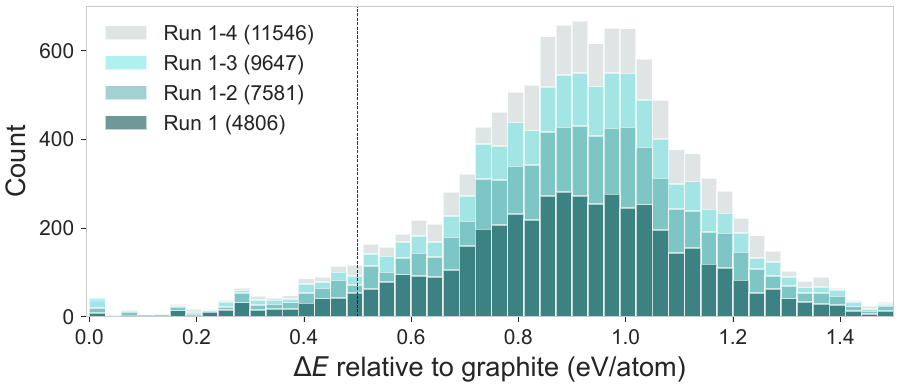} 
  \end{minipage}
  \caption{\textbf{Iterative refinement performance and energy distributions.} Left: Cumulative metrics showing progressive improvement in valid structures ($\mathrm{CN}=3$), unique structures, and low-energy candidates. Right: MACE energy distributions showing cumulative growth of validated structures across refinement runs.}
  \label{fig:multi_opt}
\end{figure*}

\subsection{The Performance of Dual-Level Refinement Strategy}
Next, we focus on improving the cumulative counts of valid, unique, and low-energy candidates by introducing latent-space refinement. In the previous SciPy--CPU workflow, processing a pool of $100$k candidates required roughly $9$--$10$~h, yet many samples still failed to reach the target local environment ($\mathrm{CN} = 3$) under their fixed symmetry conditions. This is because that a random structure drawn from $z$ can lie in an unfavorable basin of the SO(3) loss landscape for the given condition. Although the optimization in the representation space $\mathbf{R}$ can reduce the descriptor loss but still get trapped in the local minima. To escape from these unfavorable minima, we alter the optimization on $z$ to move the decoded structure into a basin closer to the target environment, and then we re-apply representation-space optimization to fine-tune the free variables. We repeat this dual-level recovery on the remaining invalid set for additional rounds and report cumulative results throughout (all counts are with respect to the same $100$k candidate pool). After Run~1, we obtain 15{,}076 valid structures and 4{,}806 unique structures; after the first latent-refinement + re-optimization round (Run~1--2), these increase to 29{,}837 valid structures and 7{,}581 unique structures. Repeating the procedure for three more rounds increases the cumulative totals to $48{,}308$ valid structures (about $48\%$) and $11{,}546$ unique structures (about $12\%$). The low-energy structure count increases from $412$ to $1{,}014$. The results are summarized in Fig.~\ref{fig:multi_opt}. In total, the full four-round workflow takes $8.56$~h on the GPU and yield about 2.5 times of unique low-energy structures. This is considerably more efficient than with than our earlier SciPy--CPU workflow in terms of both run time and success rate. 

\section{Conclusions}
In summary, we have advanced the \texttt{LEGO-xtal} framework for generative crystal structure design by introducing a conditional VAE architecture and a dual-level GPU-accelerated optimization strategy. The CVAE enables targeted generation of structures under fixed symmetry constraints, improving the yield of valid candidates exhibiting the desired sp$^2$ bonding motif. The GPU-batched optimization pipeline significantly accelerates descriptor-based refinement in the representation space, while latent-space optimization provides a complementary mechanism to escape local minima and recover stalled candidates. The iterative dual-level refinement strategy effectively combines these approaches, more than doubling the yield of unique and low-energy structures within a fixed computational budget. Overall, these methodological advancements notably enhance the efficiency of AI-driven crystal structure generation in our previous workflow. Further improvement on this workflow (e.g., extension to multi-component and multi-environment) will make it practical for rapid exploration of novel materials with target structural characteristics in the near future.

\begin{acknowledgments}
This research was sponsored by the U.S. Department of Energy, Office of Science, Office of Basic Energy Sciences, and the Established Program to Stimulate Competitive Research (EPSCoR) under the DOE Early Career Award No. DE-SC0024866, the UNC Charlotte's seed grant for data science, as well as European Union (ERDF), R\'{e}gion Nouvelle Aquitaine, Poitiers University, and French government programs ``Investissements d'Avenir" (EUR INTREE, reference ANR-18-EURE-0010) and PRC ANR  MagDesign and TcPredictor. The computing resources are provided by ACCESS (TG-DMR180040).
\end{acknowledgments}

\section*{Data Availability}
The GPU \texttt{LEGO\_xtal} source code, instructions, as well as scripts used to calculate the results of this study, are available in \url{https://github.com/MaterSim/LEGO-Xtal-GPU}.

\section*{Author Contributions}
Q.Z. H.X., and G.F. co-conceived the idea. Q.Z., H.X., and G.F. supervised this project. O.G.R., H.X., and Q.Z. implemented the code. All authors analyzed the results and contributed to manuscript writing.

\section*{REFERENCES}
\bibliographystyle{apsrev4-1}
\bibliography{ref}

\end{document}